\documentclass[3p,times,procedia]{elsarticle}

\usepackage{ecrc}

\usepackage{ulem}

\volume{00}

\firstpage{1}

\runauth{}

\usepackage{amssymb}

\usepackage[figuresright]{rotating}

\begin{document}

\begin{frontmatter}

\dochead{}
\title{Test of the Pauli Exclusion Principle in the VIP-2 underground experiment}

\author[label1,label2,label3]{C.~Curceanu}
\author[label1,label4]{H.~Shi\corref{cor1}}
\ead{Hexi.Shi@lnf.infn.it}
\author[label1]{S.~Bartalucci}
\author[label5]{S.~Bertolucci}
\author[label1,label4]{C.~Berucci}
\author[label1,label3]{A.M.~Bragadireanu}
\author[label4,label1]{M.~Cargnelli}
\author[label1]{A.~Clozza}
\author[label1]{L.~De Paolis}
\author[label6]{S.~Di Matteo}
\author[label7]{J.-P.~Egger}
\author[label1]{C.~Guaraldo}
\author[label1]{M.~Iliescu}
\author[label4,label1]{J.~Marton}
\author[label8]{M.~Laubenstein}
\author[label9]{E.~Milotti}
\author[label4]{A.~Pichler}
\author[label3,label1]{D.~Pietreanu}
\author[label2,label1]{K.~Piscicchia}
\author[label1]{A.~Scordo}
\author[label1,label3]{D.L.~Sirghi}
\author[label1,label3]{F.~Sirghi}
\author[label1]{L.~Sperandio}
\author[label10,label1]{O.~Vazquez Doce}
\author[label4]{E.~Widmann}
\author[label4,label1]{J.~Zmeskal}

\address[label1]{INFN, Laboratori Nazionali di Frascati, C.P. 13, Via E. Fermi 40, I-00044 Frascati(Roma), Italy,}
\address[label2]{Museo Storico della Fisica e Centro Studi e Ricerche Enrico Fermi, Piazza del Viminale 1, 00183 Roma, Italy,}
\address[label3]{IFIN-HH, Institutul National pentru Fizica si Inginerie Nucleara Horia Hulubbei, Reactorului 30, Magurele, Romania,}
\address[label4]{Stefan-Meyer-Institut f\"{u}r Subatomare Physik, Boltzmanngasse 3, 1090 Wien, Austria,}
\address[label5]{Dipartimento di Fisica e Astronomia, Universit\'{a} di Bologna, Italy,}
\address[label6]{Institut de Physique UMR CNRS-UR1 6251, Universit\'{e} de Rennes1, F-35042 Rennes, France,}
\address[label7]{Institut de Physique, Universit\'{e} de Neuch\^{a}tel, 1 rue A.-L. Breguet, CH-2000 Neuch\^{a}tel, Switzerland,}
\address[label8]{INFN, Laboratori Nazionali del Gran Sasso, S.S. 17/bis, I-67010 Assergi (AQ), Italy,}
\address[label9]{Dipartimento di Fisica, Universit\'{a} di Trieste and INFN-Sezione di Trieste, Via Valerio, 2, I-34127 Trieste, Italy,}
\address[label10]{Excellence Cluster Universe, Technische Universit\"{a}t M\"{u}nchen, Boltzmannstra\ss e 2, D-85748 Garching, Germany.}

\cortext[cor1]{Corresponding author}


\begin{abstract}
The validity of the Pauli Exclusion Principle, a building block of Quantum Mechanics, is tested for electrons.
The VIP (VIolation of Pauli exclusion principle) and its follow-up VIP-2 experiments at the Laboratori Nazionali del Gran Sasso 
search for x-rays from copper atomic transitions that are prohibited by the Pauli Exclusion Principle. 
The candidate events, if they exist, originate from the transition of a $2p$ orbit electron to the ground state which is already occupied by two electrons. 
The present limit on the probability for Pauli Exclusion Principle violation for electrons set by the VIP experiment is 4.7 $\times10^{-29}$.
We report a first result from the VIP-2 experiment improving on VIP limit, that solidifies the final goal to achieve a two orders of magnitude gain in the long run.
}
\end{abstract}
 
\begin{keyword}
Pauli Exclusion Principle \sep Quantum foundations \sep x-ray spectroscopy \sep Underground experiment \sep  Silicon Drift Detector 
\PACS 03.65.-w \sep 07.85.Fv \sep 32.30.Rj

\end{keyword}

\end{frontmatter}

\graphicspath{{./}}

\section{Introduction}
The Pauli Exclusion Principle (PEP) stating that in a system there cannot be two (or more) fermions with all quantum numbers identical, 
is a fundamental principle in physics. 
The validity of the PEP is the basis of the periodic table of elements, electric conductivity in metals, 
the degeneracy pressure which makes white dwarfs and neutron stars stable, 
as well as many other phenomena in physics, chemistry and biology. 
In quantum mechanics, the states of particles are described in terms of wave functions. 
For identical particles, with respect to their permutation, the states are necessarily either symmetric for bosons, or antisymmetric for fermions. 
This ``symmetrization postulate'' \cite{Mes62} excludes the mixing of different symmetrization groups and it is at the basis of the PEP. 
Messiah and Greenberg noted in \cite{Mes64} that this superselection rule ``does not appear as a necessary feature of the QM description of nature.''
In this context, the violation of PEP is equivalent to the violation of spin-statistics \cite{Gre00},
and experimentally to the existence of states of particles that follow statistics other than the fermionic or the bosonic ones.

Exhaustive reviews of the experimental and theoretical searches for a small violation of the PEP or the violation of spin-statistics can be found for example in \cite{Gre00} and \cite{Ell12}.
We point out, firstly, that there is no established model in quantum field theory that can include small violations of the PEP explicitly.
Secondly, although many experimental searches present limits for the violation, 
the parameters that quantify the limits are model/system dependent and are not generally comparable. 
Moreover, in order to search for states that are in a mixed symmetry, 
it is crucial to introduce into the system new states, 
among which the PEP-violating states may be found. 
Ramberg and Snow \cite{Ram90} took this argument into account, 
by running a high electric DC current through a copper conductor, 
and they searched for x-rays from transitions that are PEP-forbidden after electrons are captured by copper atoms. 
In particular, they searched for PEP-violating transitions from the $2p$ level to the $1s$ level, which is already occupied by two electrons.
Due to the shielding effect of the additional electron in the ground level, 
the energy of such abnormal transitions will deviate from the copper $K\alpha$ x-ray at 8 keV by about 300 eV \cite{Cur13}, 
which are distinguishable in precision spectroscopcal measurements. 
Since the $new$ electrons from the current are supposed to have no a-priori established symmetry with the electrons inside the copper atoms, 
the detection of the energy-shifted x-rays is an explicit indication of the violation of spin-statistics, and thus the violation of the PEP for electrons.

We want to mention that, 
one known system in which the dichotomy of fermions and bosons does not work is in the 2-dimensional condensed matter physics through the (fractional) Quantum Hall effect \cite{Pra90}.
Particles that are neither fermions nor bosons, and that may exist in electronic systems confined to two spatial dimensions have been constructed theoretically and 
investigated in the laboratory with great consistency with the theories as reviewed in \cite{Ste08}. 
The physics of this special system is exciting in itself and may provide hints to the searches for the violation of the PEP in other systems.

In section 2, we will introduce the VIP and VIP-2 experiments at Laboratori Nazionali del Gran Sasso (LNGS), 
and in section 3 the first results from the physics run of VIP-2 in 2016. The paper ends with conclusions and discussions.

\section{VIP-2 experiment}
The first experiment performed in the LNGS-INFN underground laboratory, 
the VIP experiment, used a similar method as that of the Ramberg and Snow, 
and the same definition of the parameter to represent the probability that the PEP is violated,
for a direct comparison of the experimental results. 
An improvement in sensitivity was achieved 
firstly by performing the experiment in the low radioactivity laboratory at LNGS, 
which has the advantage of the excellent shielding against cosmic rays. 
Secondly, the application of Charge Coupled Device (CCD) as the x-ray detector with a typical energy resolution of 320 eV at 8 keV, 
increased the precision in the definition of the region of interest to search for anomalous x-rays. 
The VIP experiment set the limit for the probability of the PEP violation for electrons to be 4.7 $\times 10^{-29}$ \cite{Cur11} \cite{Bar09} \cite{Spe08}.

\begin{figure}[htbp]
\centering
\includegraphics[width=16cm,clip]{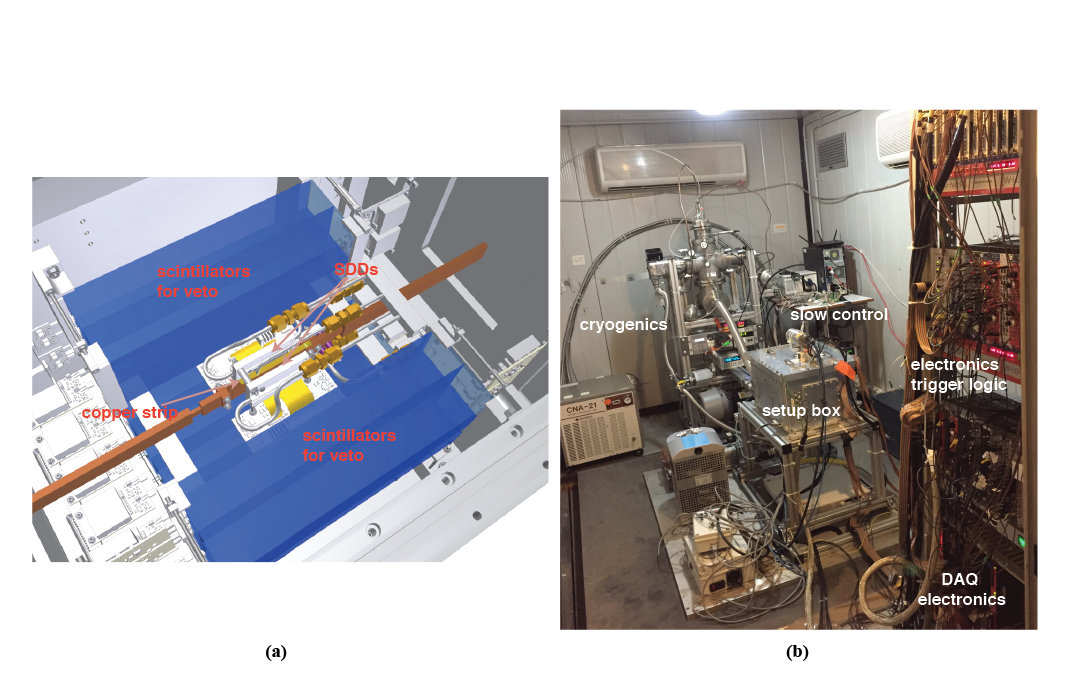}
\caption{(\textbf{a}) The design of the core components of the VIP2 setup, including the SDDs as the x-ray detector, the scintillators as active shielding with silicon photomultiplier readout;
         (\textbf{b}) a picture of the VIP2 setup in operation at the underground laboratory of Gran Sasso.
}
\label{fig:setup}
\end{figure}

By using new x-ray detectors and an active shielding of scintillators, 
the VIP-2 experiment plans to further improve the sensitivity by two orders of magnitude. 
The major improvements come from the change of the layout of the copper strip target and of the x-ray detectors, 
which allow a larger acceptance for the x-ray detection. 
Secondly a DC current with 100 Ampere is applied instead of 40 Ampere, which introduces twice more new electrons into the copper strip. 
Last but not least, in addition to the improved passive shielding surrounding the setup to reduce the background generated by the environmental radiations, 
the use of Silicon Drift Detector (SDD) as the x-ray detectors allows to implement an active shielding using scintillators, as illustrated in Figure \ref{fig:setup} (a), 
which removes the background induced by the high energy charged particles that are not shielded. 
More details of the detectors and the VIP-2 setup are given in \cite{Shi16} \cite{Pic16} \cite{Shi15}\cite{Mar13}.

The VIP-2 trigger logic was implemented using the NIM standard modules, 
and it is defined by either an event at any SDD or a coincidence between two layers of the veto detector. 
A VME-based data acquisition system for the detectors was constructed. 
It records the energy deposit of the six SDDs, from the output of a CAEN 568 spectroscopy amplifier which processes the analog signals of the SDD preamplifier output. 
The charge to digital signals (QDC) of the 32 scintillator channels, and the timing information of the SDDs with respect to the main trigger are recorded in the data as well. 
The data acquisition system can be remotely accessed and controlled from the computer terminals outside the Gran Sasso laboratory.

The temperatures of the SDDs, the copper conductor, the cooling system, and the ambient temperature, the vacuum pressure of the setup, 
are monitored by a slow control system. 
The slow control which can be accessed from remote terminals also controls the DC power supply to switch on and off the current applied to the copper strip. 
A closed circuit chiller coupled to a cooling pad attached to the copper strips keeps a constant temperature below 25 Celsius of the strips 
when the DC current up to 100 Ampere is applied. 
The temperature of the SDDs' holder frame had a change of less than 2 Kelvin when the 100 Ampere current is applied to the copper strip. 
At this level of temperature variation, the effect of change in the energy resolution of the SDDs is negligible.

In November 2015, after having performed exhaustive tests in the laboratory, 
the VIP-2 setup was transported and mounted in the Gran Sasso underground laboratory as shown in Figure \ref{fig:setup} (b).
After tuning and optimization, from October 2016 we started the first campaign of data taking with the complete detector system. 
The energy calibration of the SDDs was performed in in-situ, 
by placing near the detectors a weak Iron-55 source covered by a 25 um thick titanium foil.
The manganese $K$-series x-rays from the source partly go through the foil and partly irradiate the foil generating titanium $K$-series x-rays. 
These fluoresence x-rays are detected by the SDDs at an overall rate of about 2 Hz, and provide reference energy peaks to calibrate the digitized SDD signals to energy scale.

\section{First VIP-2 results}

\begin{figure}[htbp]
\centering
\includegraphics[width=12cm,clip]{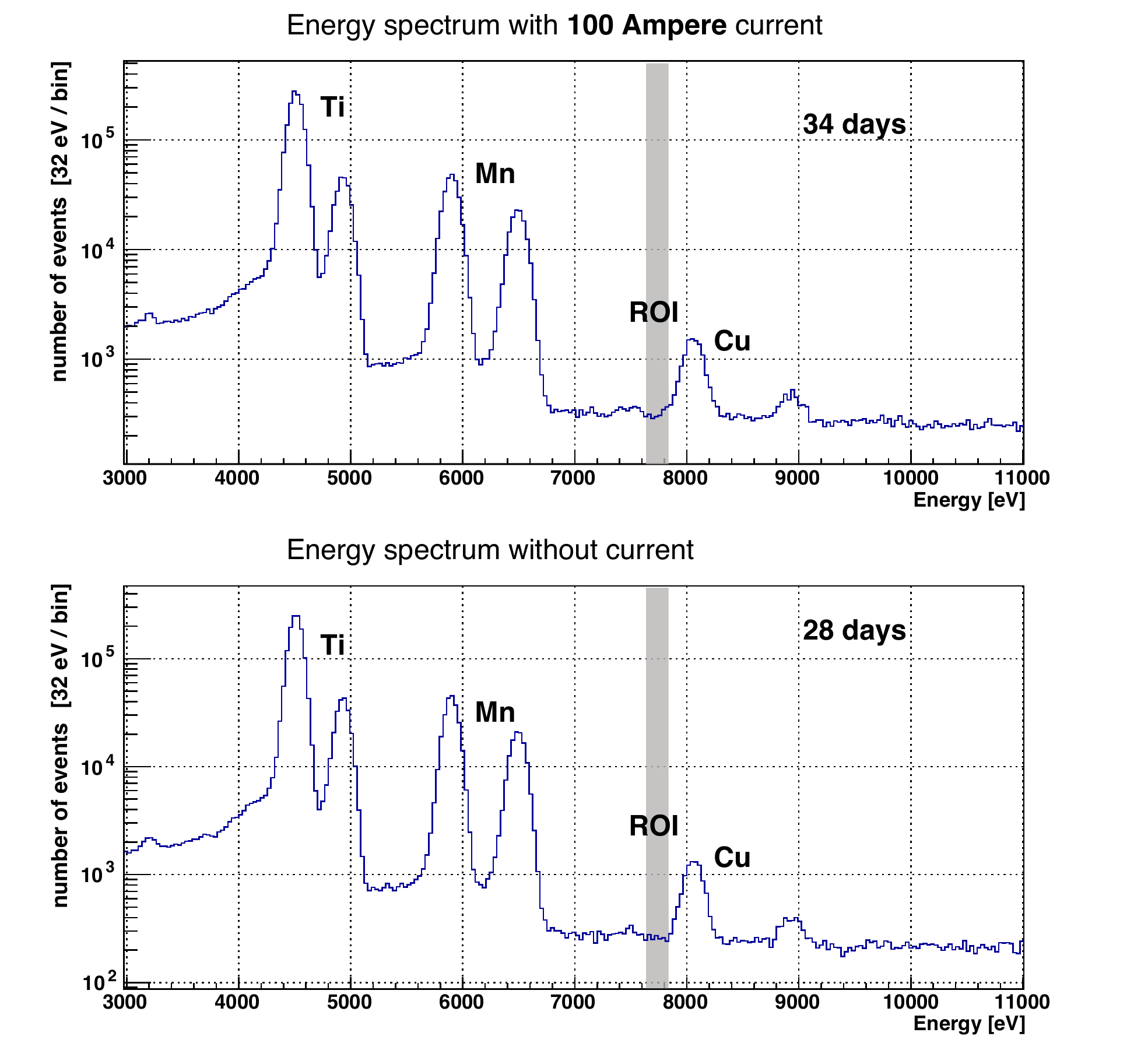}
\caption{The energy spectra from all the SDDs, for data with and without applied DC current to the copper strip, 
         taken during the physics run in late 2016 at the LNGS. 
}
\label{fig:spectra}
\end{figure}

During the data taking from October to December 2016, the DC current was typically switched on for one week and off for the next. 
The energy calibrations for the SDDs were done for each data set corresponding to a period of about one week, 
and then summed separately over the whole data taking period of over two months, for 100 Ampere current-on data and current-off data sets. 
The spectra that correspond to 34 days of effective data acquisition with 100 Ampere current on and 28 days with current off are shown in Figure \ref{fig:spectra}, 
in which the fluorecence lines of titanium and manganese are marked. 

The environmental gamma radiations and high energy charged particles can irridiate the copper conductor or the strip inside the setup, 
and the normal $K$-series x-rays from the de-excitation of the copper form the main background near the energy region of interest (ROI in Figure \ref{fig:spectra}) from 7629 eV to 7829 eV, 
which is defined by the SDD energy resolution (200 eV FWHM) at the $K_{\alpha}$ copper transition (8.04 keV) near the expected value of the PEP violating transition. 
In order to obtain the number of events violating PEP in the ROI, the current-off spectrum was normalized to 34 days of data taking time, 
and then a subtraction with the current-on spectrum was performed. 
The numbers of x-rays in the region of interest were :
\begin{itemize}
\item	with I = 100 A;  $N_X$ = 2222 $\pm$ 47 (for 34 days of data taking );
\item   with I = 0 A; \quad   $N_X$ = 2181 $\pm$ 47 (28 days of data taking normalized to 34 days);
\item	numerical subtraction : $\Delta N_X$ = 41 $\pm$ 66 (normalized to 34 days of data taking time). 
\end{itemize}

Following the similar notations used by Ramberg - Snow and the VIP experiment papers, 
the number of possible PEP violating events, $\Delta N_X$, is related to the $\beta ^2$/2 parameter giving the probability of PEP violation \cite{Gre87} : 
\begin{eqnarray}
\Delta N_X &\ge& \frac12 \beta^2 N_{new}\frac{1}{10} N_{int} \times ({\rm detection\quad efficiency \quad factor}) \nonumber \\
&=&  \frac{\beta^2(\Sigma I\Delta t)D}{e\mu}\frac{1}{20} \times ({\rm detection\quad efficiency \quad factor}). 
\end{eqnarray}

Furthermore, the number of new electrons that pass through the conductor, 
\begin{equation}
N_{new} = (1/e)\Sigma I \Delta t, 
\end{equation}
is given by the electric charge $e$ of the electron, the intensity $I$ of the applied DC current, 
and the duration time $\Delta t$ of the measurement. 
The minimum number of internal scattering processes between a new electron and the atoms of the copper lattice, 
$N_{int}$, is of order $D/\mu$, where $D$ is the length of the copper strip (10 cm), and $\mu$ is the mean free path of electrons in copper. 
We follow the same assumption used in the VIP paper \cite{Bar06}, 
that the capture probability of a new electron by an atom of the copper lattice is greater than 1/10 of the scattering probability. 

The detection efficiency factor, is evaluated with a Monte Carlo simulation based on Geant4.10 with realistic detector configuration, taking into account : 
the transmission rate of a copper $K_{\alpha}$ x-ray that origins at a random position inside the copper strip and reaches the surface; 
the geometrical acceptance of the photons coming from the surface of the copper stipe arriving at the six SDD detectors; 
the detection efficiency of a copper $K_{\alpha}$ x-ray by the 450 um thick SDD unit, 
and the value is determined to be about 1\%.

With $D$ = 10 cm, $\mu$ = 3.9 $\times$ 10$^{-6}$ cm, $e$ = 1.602 $\times$ 10$^{-19}$ C, $I$ = 100 A, 
and normalizing the measurement time with current to 34 days, 
using the three sigma upper bound of $\Delta N_X$ = 41 $\pm$ 66 to give a 99.7\% C.L., 
we get an upper limit for the $\beta^2$/2 parameter : 
\begin{equation}
\frac{\beta^2}{2} \le \frac{3\times 66}{4.7\times 10^{30}} = 4.2 \times 10^{-29}.
\end{equation}

\begin{figure}[htbp]
\centering
\includegraphics[width=9 cm,clip]{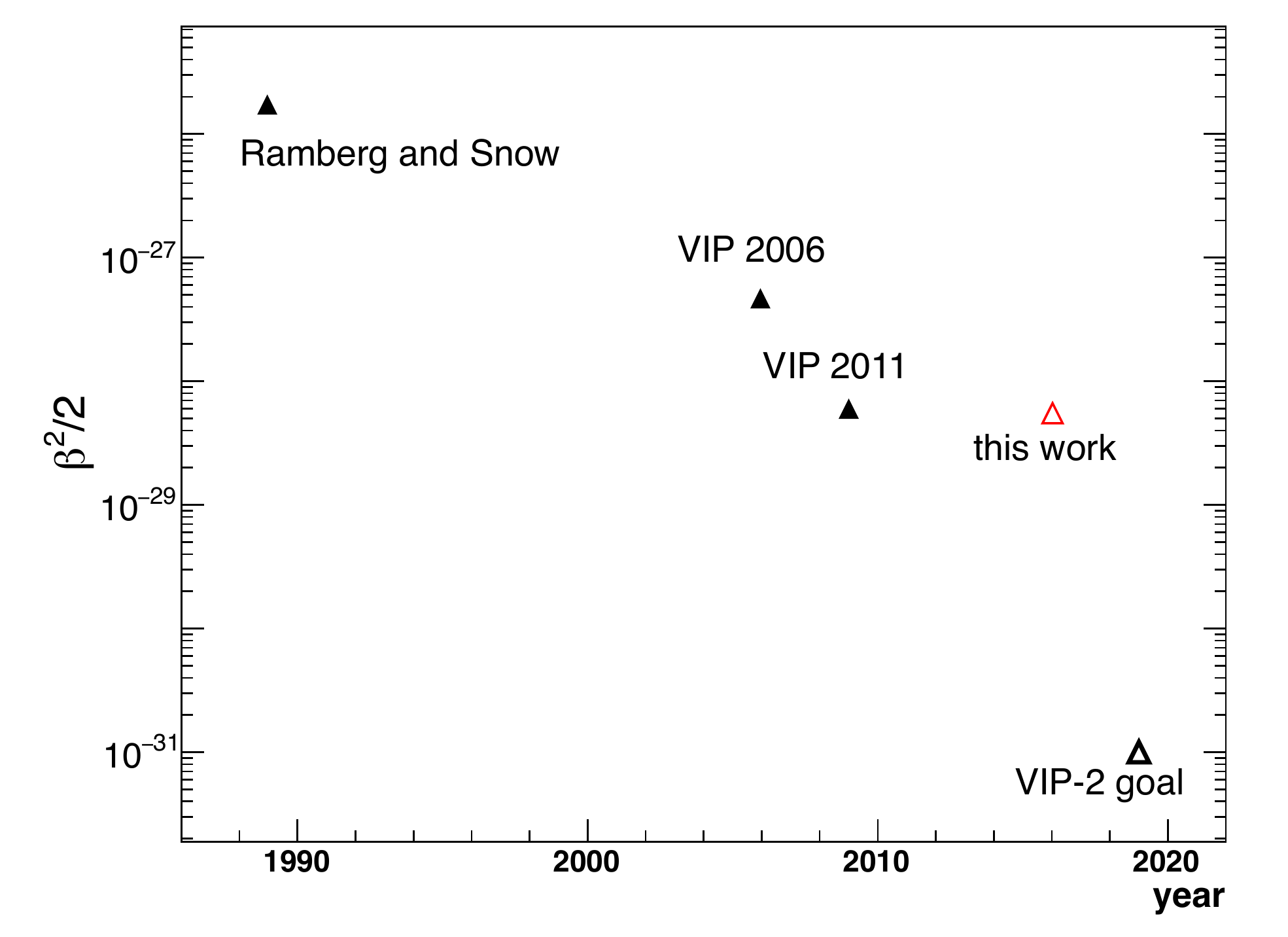}
\caption{ All the past results from PEP violation tests for electrons with a copper conductor, 
          together with the result from this work and the anticipated goal of VIP-2 experiment. 
          Note that the result of this work comes from two months of data taking, 
          and it is already compatible with the VIP result from three years of operation.
}
\label{fig:betaSquare}
\end{figure}

\section{Conclusions and Future Perspectives}
The first VIP-2 physics run from two months of data taking already gave a better limit than the VIP result obtained from three years of running.

In Figure \ref{fig:betaSquare}, we show all the past experimental results of the PEP violation tests for electrons with a copper conductor, together with this work. 
The new result shows that in planned data taking time of 3 to 4 years, 
the VIP-2 experiment can either set a new upper limit for the probability that the PEP is violated at the level of $10^{-31}$, 
improving the VIP experiment result by two orders of magnitude, 
or find the PEP violation, which would have profound implications in science and philosophy.

We conclude with the words of Lev Okun from his 1987 paper \cite{Oku87}: 
``{\it The special place enjoyed by the Pauli principle in modern theoretical physics does not mean that this principle does not require further and exhaustive experimental tests. 
On the contrary, it is specifically the fundamental nature of the Pauli principle which would make such tests, 
over the entire periodic table, of special interest.}''

\vspace{6pt}

\paragraph{Acknowledgements}
We thank H. Schneider, L. Stohwasser, and D. St\"{u}ckler from Stefan-Meyer-Institut 
for their fundamental contribution in designing and building the VIP2 setup. 
We acknowledge the very important assistance of the INFN-LNGS laboratory staff during all phases of preparation, installation and data taking.
We thank the Austrian Science Foundation (FWF) which supports the VIP2 project with the grant P25529-N20. 
We acknowledge the support from the EU COST Action CA15220, 
and from Centro Fermi (``Problemi aperti nella meccania quantistica''project).
Furthermore, this paper was made possible through the support of a grant from the John Templeton Foundation (ID 58158).
The opinions expressed in this publication are those of the authors and do not necessarily reflect the views of the John Templeton Foundation.


\bibliographystyle{elsarticle-num}

\end{document}